\documentclass[nofootinbib,superscriptaddress,twocolumn]{revtex4-1}
\usepackage{wrapfig} 
\usepackage{graphicx} 
\usepackage{float}
\usepackage{subfigure}
\usepackage{amssymb} % for, e.g., real numbers symbol

\usepackage{color}

\newcommand{\ket}[1]{\left|#1\right\rangle}
\newcommand{\bra}[1]{\left\langle #1\right|}

\usepackage{amsmath}
\DeclareMathOperator*{\argmax}{argmax}

\usepackage[dvipsnames]{xcolor}
\usepackage[colorlinks=true,linkcolor=blue,citecolor=blue]{hyperref}

\makeatother

\begin{document}
\title{
Predicting quantum advantage by quantum walk with convolutional neural networks 
}
% Detecting quantum walk speedup with convolutional neural networks

\author{Alexey A. Melnikov}
\thanks{Corresponding author, e-mail: alexey.melnikov@unibas.ch, \\website: \href{https://melnikov.info}{melnikov.info}}
\affiliation{ITMO University, Kronverksky prospekt 49, 197101 St. Petersburg, Russia}
\affiliation{Valiev Institute of Physics and Technology, Russian Academy of Sciences, Nakhimovskii prospekt 36/1, 117218 Moscow, Russia}

\author{Leonid E. Fedichkin}
\affiliation{Valiev Institute of Physics and Technology, Russian Academy of Sciences, Nakhimovskii prospekt 36/1, 117218 Moscow, Russia}
%\affiliation{NIX, Zvezdny boulevard 19, 129085 Moscow, Russia}
\affiliation{Moscow Institute of Physics and Technology, Institutskii pereulok 9, 141700 Dolgoprudny, Moscow Region, Russia}

\author{Alexander Alodjants}
\affiliation{ITMO University, Kronverksky prospekt 49, 197101 St. Petersburg, Russia}

\begin{abstract}
Quantum walks are at the heart of modern quantum technologies. They allow to deal with quantum transport phenomena and are an advanced tool for constructing novel quantum algorithms. Quantum walks on graphs are fundamentally different from classical random walks analogs, in particular, they walk faster than classical ones on certain graphs, enabling in these cases quantum algorithmic applications and quantum-enhanced energy transfer. However, little is known about the possible advantages on arbitrary graphs not having explicit symmetries. For these graphs one would need to perform simulations of classical and quantum walk dynamics to check if the speedup occurs, which could take a long computational time. Here we present a new approach for the solution of the quantum speedup problem, which is based on a machine learning algorithm that predicts the quantum advantage by just ``looking'' at a graph. The convolutional neural network, which we designed specifically to learn from graphs, observes simulated examples and learns complex features of graphs that lead to a quantum advantage, allowing to identify graphs that exhibit quantum advantage without performing any quantum walk or random walk simulations. The performance of our approach is evaluated for line and random graphs, where classification was always better than random guess even for the most challenging cases. Our findings pave the way to an automated elaboration of novel large-scale quantum circuits utilizing quantum walk based algorithms, and to simulating high-efficiency energy transfer in biophotonics and material science.

\end{abstract}

\maketitle

%%% speedup
Computational speedup is one of the keystone problems both in classical and quantum computer sciences~\cite{Montanaro2016,Lidar2014}. Although quantum parallelism, in general, represents necessary ingredient for an acceleration of computational algorithms on quantum ``hardware'', sufficient criterion is still unknown in many cases. Strictly speaking, speedup problem might be recognized for certain computational tasks for which definite classical and/or quantum algorithms are used, cf.~\cite{Cirac2018,Boixo2018}. In the paper we attack the speedup problem with random and quantum walks in a quite general form by using advanced machine learning approaches.

%%% about quantum and random walks
Random walks on graphs are widely used as subroutines in computational algorithms~\cite{rajeev1995randomized,landau2001efficient,szummer2002partially,grady2006random,gkantsidis2006random}, and as a model for processes in nature~\cite{kac1947random,sottinen2001fractional,bartumeus2005animal,brockmann2006scaling,codling2008random}. Quantum walks~\cite{PhysRevA.48.1687,doi:10.1080/00107151031000110776,venegas2008quantum,venegas2012quantum}, quantum analogs of classical walks, replace a classical particle with a quantum one. This change makes a fundamental difference in the walker's dynamics due to quantum interference. Due to interference, quantum walks of single and multiple particles can be employed as a tool for quantum information processing and quantum algorithms~\cite{doi:10.1142/S0219749903000383,Ambainis:2007:QWA:1328722.1328730,Childs1,doi:10.1137/090745854,Childs2,portugal2013quantum,melnikov2016quantum,chakraborty1016spatial}, for quantum machine learning~\cite{briegel2012projective,paparo2014quantum}, and as a part of a model of photosynthetic energy transfer~\cite{engel2007evidence,mohseni2008environment}. Especially in the energy transport problem, see, e.g.,~\cite{scholak2011efficient,manzano2012quantum,asadian2013heat}, it is important that quantum particles hit target vertices of certain graphs faster than classical particles.

%%% about different approaches in predicting the speedup and their drawbacks
It is, however, not clear on which graphs and for which target vertices quantum advantage will be present. Given a graph, a standard approach would be to simulate quantum and classical dynamics, and observe in which case a particle would reach a target vertex faster. However, this approach has several limitations. First, although the propagation time scales polynomially in the size of the graph~\cite{lawler86expected,lovsz2002random}, the simulations can become computationally difficult for large graphs. Second, one is usually interested in a set of graphs rather than in a single graph, which makes the simulations limited by the set size. Third, results of the simulations do not reveal any pattern, or general laws, of quantum advantage in stochastic propagation. Given the results of the simulations on one set of graphs, it is not clear how particles propagate on another set of graphs. To overcome these limitations, one can follow a different approach of theoretically investigating the walk dynamics and obtaining analytical results. This was done, e.g., for line~\cite{ambainis2001one}, cycle~\cite{aharonov2001quantum,solenov2006continuous,fedichkin2006mixing}, hypercube~\cite{kempe2005discrete,PhysRevA.73.032341}, complete~\cite{SANTOS_QUANTUM_2010}, and glued trees graphs~\cite{Childs:2003:EAS:780542.780552}, for certain target vertices. Knowing that one of these graph types is embedded in a larger graph structure can also give information about possible quantum speedups~\cite{makmal2014quantum,makmal2016quantum}. This analytical approach is, however, limited to known cases, which is only a tiny subset of all labeled connected graphs~\cite{slone1964online}.

%%% transition to our work
Combining the two described approaches to predict quantum advantage is potentially possible. An expert, looking at a graph, might recognize a known structure (e.g., a hypercube), and with a help of a limited number of simulations, draw a conclusion about a possibility of a quantum speedup. But an expert cannot possibly analyze a large number of graphs. Could a machine be able to do that and predict quantum speedups? We answer this question in the affirmative and extend the list of new machine learning techniques successfully applied in physics~\cite{biamonte2017quantum,dunjko2018machine,carleo2017solving,carrasquilla2017machine,chng2017machine,melnikov2018active,bukov2018reinforcement,fosel2018reinforcement,odriscoll2019hybrid,nautrup2018optimizing,iten2018discovering,wallnofer2019machine}.

%%% about machine learning approach to that problem: our approach, CNN intro
In this paper we take a supervised learning approach to predict a quantum speedup: a convolutional neural network (CNN)~\cite{lecun1998gradient,lecun2015deep} learns from examples to recognize quantum speedup. CNNs are widely used for image classification~\cite{krizhevsky2012imagenet}, visual document analysis~\cite{simard2003best}, face recognition~\cite{lawrence1997face}, and video classification~\cite{karpathy2014large}. Here we use a CNN for graph features extraction and learning the most relevant features, which we apply to a classification problem defined within the quantum walk framework. CNNs were recently used with graph adjacently matrix input for predicting clinical neurodevelopmental outcomes from brain networks~\cite{kawahara2017brainnetcnn} and for classifying and predicting the presence of super-diffusion in multiplex networks~\cite{leli2018deep}.

%%% Our results
Our results of using the CNN of special architecture, which we call classical-quantum convolutional neural network (CQCNN), demonstrate that the network can to represent the quantum speedup by quantum walk. CQCNN is able to generalize and correctly predict quantum speedup for unseen line graphs and random graphs with up to $25$ vertices. The quantitative classification results differ depending on the type of a graph, on the type and quantity of training examples, and on the number of training epochs. Independent of the scenario and the difficulty, however, we observe that the CQCNN is better than a random guess. Importantly, we show that it is possible to extract the logic behind a classifier function constructed by the neural network, which lets us understand and verify how the classification works on small graphs.

%%% Importance of our results
We believe that the proposed learning model will be of a particular significance for physical implementations of quantum-enhanced transport systems. A physical implementation of quantum walks is not unique: it depends on measurement procedure properties, as well as on particular properties of the physical system including experimental imperfections\footnote{Current photonic technologies~\cite{fulvio2018photonic} represent versatile platform for experimental studies of bosonic quantum walks and speedup prediction on the graphs with desirable sizes and topology~\cite{Walther2012,Szameit2018}. Moreover, a design of CMOS-compatible large scale quantum photonic devices gives hope to a realization of quantum walks based algorithms in nearest future~\cite{Brein2018}.}. In the case of such a quantum experiment, only a limited number of data points can be realistically obtained, which will make the proposed autonomous learning algorithm essential for successful implementation of the quantum-enhanced transport systems.

\section*{Results}

\begin{figure}[!t]
	\centering
	\includegraphics[width=1\linewidth]{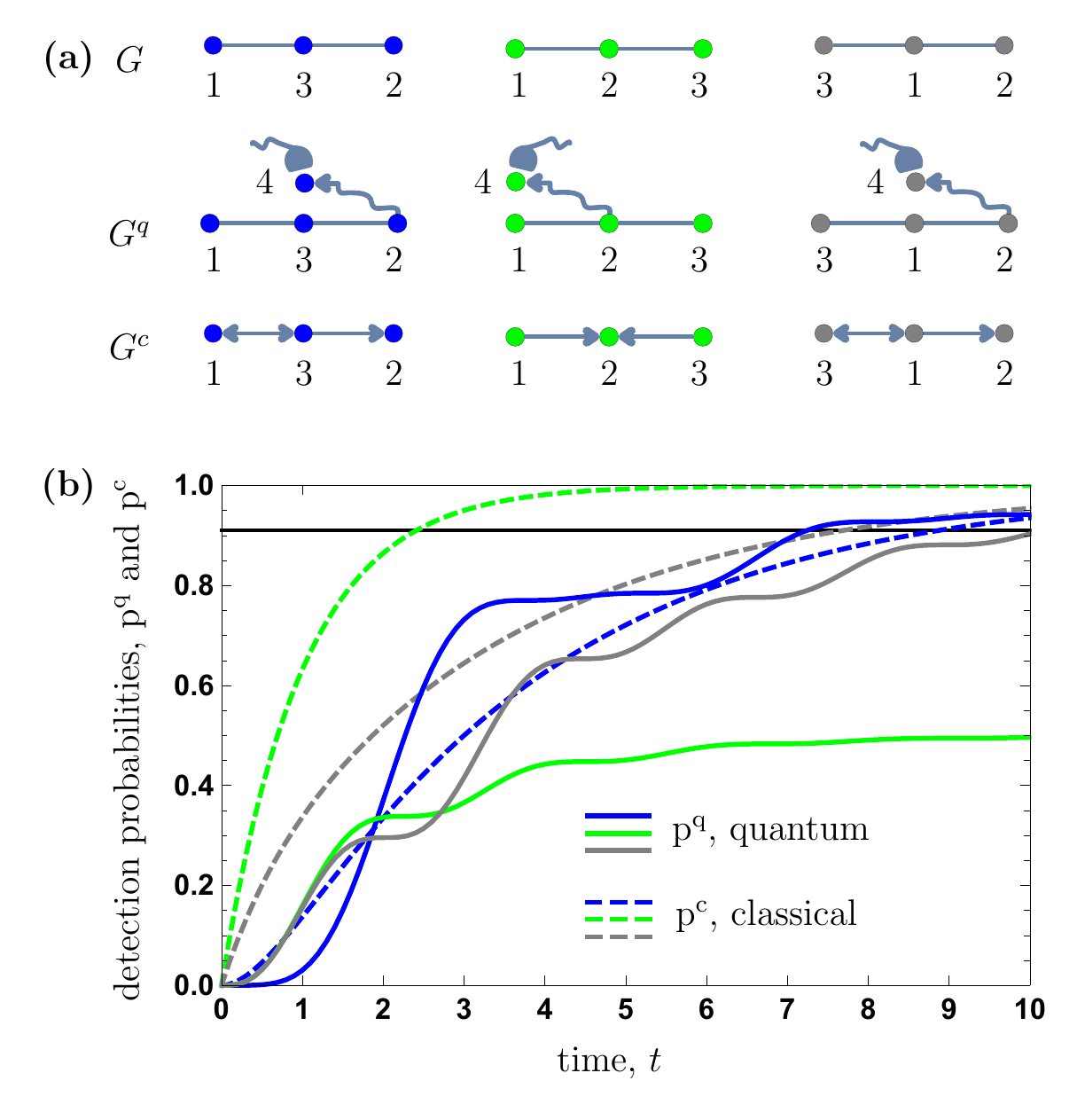}
	\caption{Quantum and classical walks on line graphs. $``1"$ is the initial vertex, whereas $``2"$ is the target vertex. (a) Inequivalent line graphs with three vertices are depicted in three different colors (blue, green, and gray). The graphs $G^q$ and $G^c$ graphs are modifications of the graphs $G$ that take into account different aspects of the physical implementation of quantum and classical walks, respectively. (b) The quantum (solid) and the classical (dashed) walk dynamics on three different line graphs are shown. The black line at the value of $1/\log 3\approx 0.91$ is the probability threshold at which particle is considered to be detected.}
	\label{fig:Lines3}
\end{figure}

\begin{figure*}[!ht]
	\centering
	\includegraphics[width=1\linewidth]{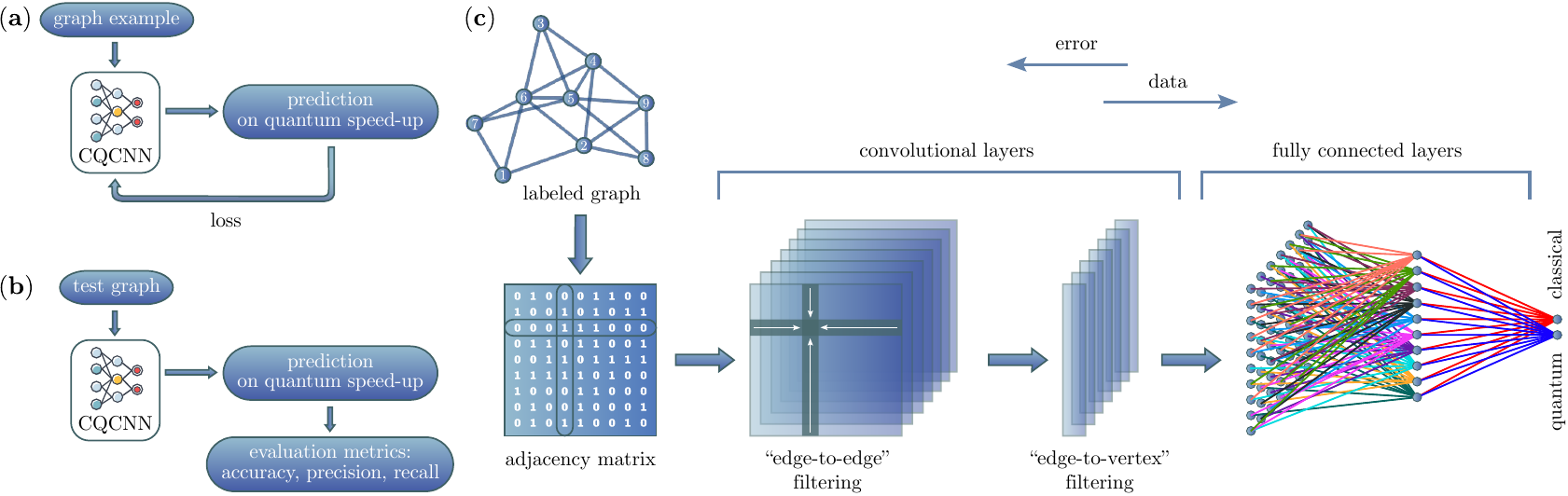}
	\caption{The machine learning approach that is used for predicting the quantum speedup. (a) A process of training CQCNN. (b) A process of testing CQCNN. (c) A scheme of the CQCNN architecture. The neural network takes a labeled graph in form of an adjacency matrix as an input. This input is then processed by convolutional layers with graph-specific ``edge-to-edge" and ``edge-to-vertex" filters (see Methods). The convolutional layers are connected with fully-connected layers that finally classify the input graph. The number of layers is the same for all graph sizes. Data and error propagation are shown with arrows.}
	\label{fig:schemes}
\end{figure*}

% A scheme of CQCNN that is used to classify quantum and classical graph properties. The network consists of an input layer, several convolution layers, two fully connected layers with nonlinear activation functions, and two output neurons. Feature maps are shown as squares and columns. See main text for details.

Quantum and classical random walk processes have different dynamics, which leads to a difference in how fast particles traverse graphs from an initial vertex to a target vertex. This difference depends not only on the nature of the particles, but also on the graph on which the particles walk. Importantly, the graph is specified not only by the way vertices are connected, but also by the positions of the initial and the target vertices. It is known that, e.g., quantum particles on line graphs reach target vertices on distance $d$ quadratically faster in $d$~\cite{ambainis2001one}. But if initial and target vertex are not far from each other, it is not easy to determine which particle is faster. To give an instructive example, let us consider line graphs, as random walks on lines are one of the simplest and most extensively studied stochastic processes~\cite{rajeev1995randomized}. In the case of three vertices, there are three inequivalent graphs $G$ shown in the first row of Fig.~\ref{fig:Lines3}(a). Complementary to graphs $G$, two additional rows of graphs are depicted: $G^q$ and $G^c$. These graphs are modifications of $G$, and correspond to the physical implementation of $G$ for quantum ($G^q$) and classical ($G^c$) walks. In the classical case, the target vertex is connected to the neighboring vertices by directed edges. In the quantum case, the sink vertex $4$ connected to the target vertex is used to measure the quantum particle, the rest of the graph is unchanged. The measurement process hence changes the dynamics of the quantum system.

Figure~\ref{fig:Lines3}(b) represents the results on quantum (solid lines) and classical (dashed lines) random walk simulations for all three graphs (blue, green, and gray). We can see that in two cases the classical walker is faster than the quantum one (green and gray cases), and the quantum particle is faster in one case (blue). From this toy example it is clear that the quantum transport speedup is only present in case of the initial and the target vertices being on opposite sites of the graph; and the classical particles are faster if these two vertices are directly connected.

We next describe how the neural network, CQCNN, can learn this for larger graphs and show the results of the learning processes. The learning setup that we use in the paper is depicted in Fig.~\ref{fig:schemes}. Fig.~\ref{fig:schemes}(a) shows schematically how CQCNN is trained using examples of graphs. CQCNN at each step takes a graph as an input in the form of an adjacency matrix, and outputs a prediction about the class this graph belongs to (quantum or classical). Having a correct label, the loss value is computed. Fig.~\ref{fig:schemes}(b) depicts the testing procedure. The difference from the training process is that CQCNN does not receive any feedback on its prediction. In the testing process the network is not modified. The neural network architecture is shown in Fig.~\ref{fig:schemes}(c). CQCNN has a layout with convolutional and fully connected layers, and two output neurons that specify two possible output classes. The convolutional layers are used to extract features from graphs, and decrease the dimensionality of the input. By trying different approaches, we observed that relevant features are not in the small local blocks of the adjacency matrices, but in the rows and columns of these matrices. We therefore constructed filters in the form of ``crosses'' shown in Fig.~\ref{fig:schemes}(c) to capture a weighted sum of column and row elements. These filters act as functions of a weighted total number of neighboring vertices of each vertex. As we will show next, the cross ``edge-to-edge'' and ``edge-to-vertex'' filters demonstrate that the convolutional network can predict the quantum advantage by quantum walk.

%\\\medskip
%\\\bigskip\bigskip\bigskip
\subsection*{Predicting quantum advantage for line graphs}
% Detecting quantum advantage for line graphs
% Machine learning to represent quantum advantage
% Machine learning to detect the quantum advantage
% Convolutional neural networks for learning graph properties
% Learning to detect a quantum speedup with convolutional neural networks

\begin{figure*}[ht!]
	\centering
	\includegraphics[width=1\linewidth]{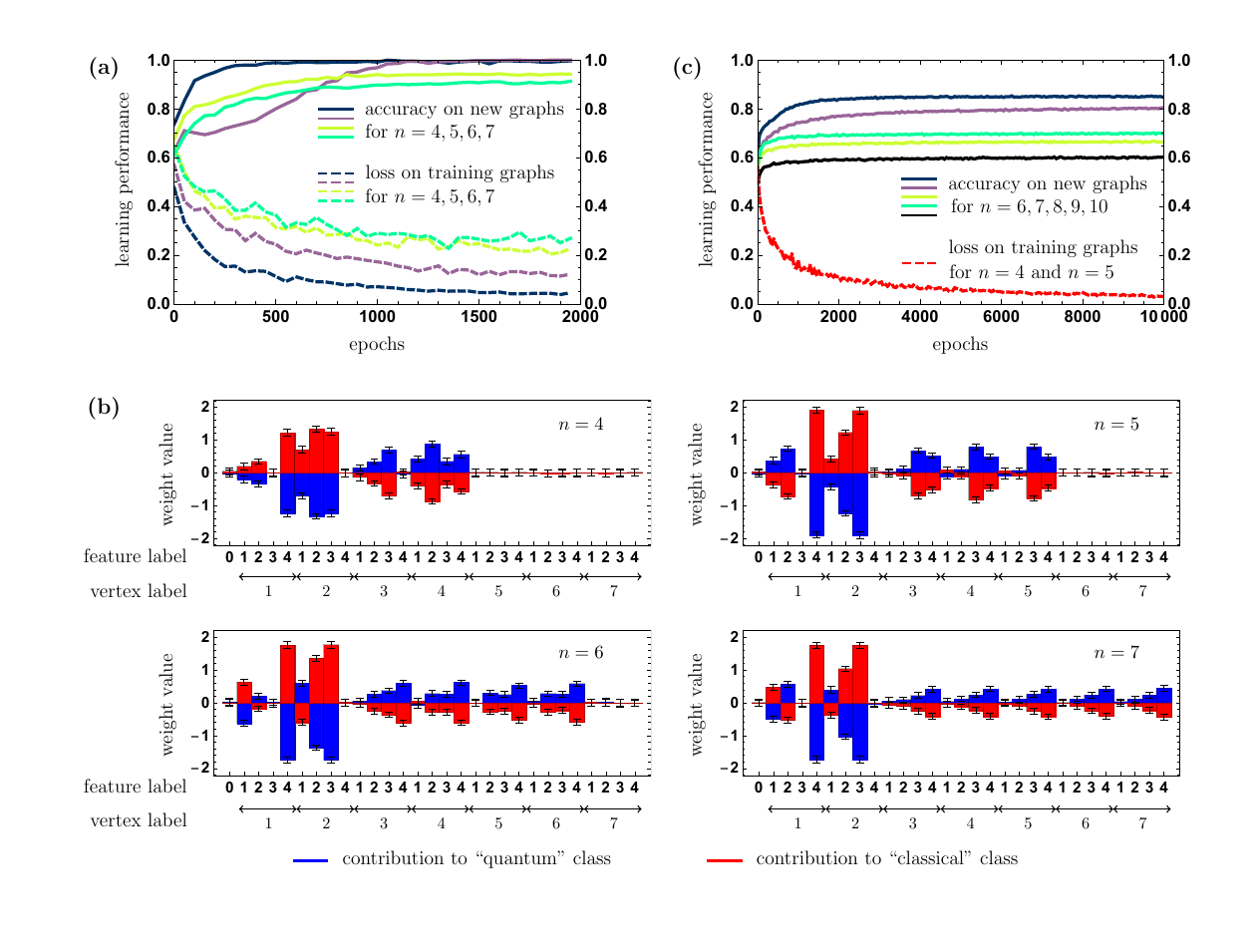}
	\caption{(a) Learning performance of CQCNN. The dataset consists of line graphs with $n=4,5,6$ and $7$ vertices, and the corresponding classical and quantum labels. These results are the average over $100$ independent CQCNNs. (b) The average values of CQCNN weights in the last layer are shown as bars. Blue and red bars correspond to the weights, shown in blue and red in Fig.~\ref{fig:schemes}(c), that connect feature maps to the ``quantum'' class and the ``classical'' class, respectively. The results are the average over $100$ independent CQCNNs. Mean squared deviation is shown as a vertical line for each bar. (c) Learning performance of CQCNN that is trained on graphs with $4$ and $5$ vertices, and tested on graphs with $6-10$ vertices.}
	\label{fig:linesLearning}
\end{figure*}

We apply the described machine learning methodology to different sets of graphs. In order to understand how our approach works in a systematic way, we first analyze the neural network performance on line graphs. We take the simplest design of CQCNN in Fig.~\ref{fig:schemes} and apply it to line graphs with up to $10$ vertices. We trained CQCNN over $2000$ epochs with a single batch of $3$ examples per epoch. The results of these simulations are shown in Fig.~\ref{fig:linesLearning}. Eight lines of four different colors in Fig.~\ref{fig:linesLearning}(a) demonstrate the results of training the neural network on line graphs; each color corresponds to a specific size of a graph with $n=4,5,6,7$ vertices. For the simulations we used datasets with all possible line graph labeling: $90\%$ of which is used to train (dashed lines) CQCNN, and $10\%$ are used to test (solid lines) its generalization capabilities. The performance of CQCNN on the training graphs is defined by the cross entropy loss function. The loss on a test example $i$ is defined relative to the correct class $\mathrm{class}_i$ (classical or quantum, $0$ or $1$) of this example:
\begin{equation}\label{lossFunction}
\mathrm{loss}_i = - \kappa(\mathrm{class}_i)\log \left(\frac{\mathrm{e}^{x(\mathrm{class}_i)}}{\mathrm{e}^{x(0)}+ \mathrm{e}^{x(1)}}\right),
\end{equation}
where $\kappa(\mathrm{class}_i)$ is the total fraction of examples from this class in the dataset, $x(0)$ and $x(1)$ are the values of the output neurons. In Fig.~\ref{fig:linesLearning}(a) one can see that CQCNN learns to represent the training graphs as the loss defined by Eq.~(\ref{lossFunction}) goes down (dashed curves). But most importantly CQCNN constructed a function that generalizes over seen graphs to unseen graphs, as the classification accuracy\footnote{Classification accuracy is the fraction of correct predictions.} goes up (solid curves).

\begin{figure*}[ht!]
	\centering
	\includegraphics[width=1\linewidth]{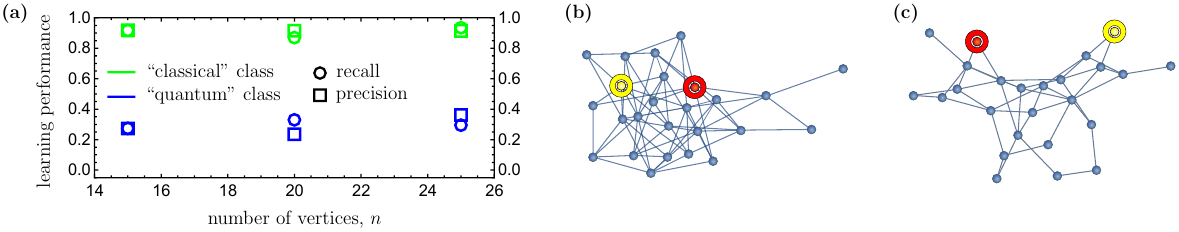}
	\caption{(a) Learning performance of CQCNN. The dataset consists of random graphs with $n=15, 20$ and $25$ vertices, $1000$ examples for each $n$, and the corresponding classical and quantum labels. CQCNN was simulated during $3000$ epochs, $100$ mini batches each with the batch size of $3$ examples. The neural network was tested on $1000$ random graphs for each $n$. (b)-(c) An example of two graphs from the test set which were correctly classified by the neural network. On graph (b) the classical particle is faster, whereas on the graph (c) the quantum particle is faster. The initial and the target vertices are marked in yellow and red, respectively.}
	\label{fig:randomLearning}
\end{figure*}

Our results in Fig.~\ref{fig:linesLearning}(a) demonstrate that it is possible for CQCNN to learn a function that maps graphs to their quantum walk properties. In order to understand the predictive capacity of CQCNN, we analyze the weights of the fully connected layer of the simple CQCNN employed for this classification problem. These weights are visualized in Fig.~\ref{fig:linesLearning}(b) as bars for different number $n=4,5,6$ and $7$ of vertices. In total, one can see $29$ bars corresponding to $29$ weights in the last layer of the neural network. These weights form a feature vector which we divide into $7$ parts, each corresponding to a specific vertex of the graph and labeled as ``vertex label'' in Fig.~\ref{fig:linesLearning}(b). Each vertex has $4$ features labeled as $1-4$, the zero-th component of the feature vector is the bias. All these features are outputs of the convolutional layers, and their values have, up to some learned coefficients, the following meaning. The first feature for each vertex corresponds to the number of edges this vertex has, the second feature -- to the total number of neighboring edges of all edges leading to the vertex. The third feature gives one if the vertex is connected to the initial vertex by an edge, and zero otherwise. The fourth feature does the same relative to the target vertex. Note that some weights are learned to be zero for vertices that are not present in smaller graphs.

By looking at the weights of CQCNN, we observe that the designed neural network learned several properties of the quantum advantage on these line graphs. 
First, we observe that the contributions to the quantum (blue) and classical (red) classes are symmetric: whatever is a positive indication of the quantum class -- it is a negative indication of the classical class. Second, the weights are different for different vertices, and this difference explains the classification outcome as we describe next. The graph shows no quantum advantage if the initial vertex is connected to the target vertex (the feature $4$ for the vertex $1$, and the feature $3$ for the vertex $2$). It is also discouraged if the target vertex is well connected to the rest of the graph (the features $1$ and $2$ for the vertex $2$). And, although the weights of the other features do not strongly define the role of these features, the more connected these vertices -- the better for the quantum speedup. 

The landscape of weights changes when the graph size grows (growing $n$ in Fig.~\ref{fig:linesLearning}(b)), but not drastically. The described correlations hold for all studied graph sizes. In addition to this consistency, we see that the deviation of weights from their average is quite small -- all $100$ CQCNNs learned very similar weights. By looking at vertices $3,4,5,6$, and $7$, we observe that the weights are almost identical: all these vertices contribute identically to the classification. Indeed, as it turns out, the dynamics of particles is invariant under relabelling of the vertices apart from the initial and the target vertices. Hence CQCNN autonomously realized that many graph examples are isomorphic.

Learning all these graph properties helps the network to correctly classify graphs of the same size which were not seen previously. CQCNN can go a step further, and apply the learned data representation to graphs of larger sizes. This can be seen in Fig.~\ref{fig:linesLearning}(c) where the training is done on line graphs with $n=4$ and $5$, but tested on graphs with $n=6-10$ vertices. The classification accuracy on larger graph sizes is between $60$ and $85\%$, which is significantly better than a random guess. Note that the generalization performance is not $1$, as we observed that for different graph sizes there are always new cases that are not derived from the smaller graphs. It is also the case that generalization performance goes down with $n$ relatively fast, hence suggesting that at least more training examples will be needed for graphs of larger size. Importantly, as proven by the results in Fig.~\ref{fig:linesLearning}(c), the CQCNN approach is flexible to a change of graph sizes without being trained on all sizes.

\subsection*{Predicting quantum advantage for random graphs}

CQCNN was shown to be able to classify line graphs. Next, we estimate how well the presented methodology works on other graphs. In general, the more symmetries the graph has -- the better we would expect CQCNN's performance is, as there are more ways to learn graph properties from examples. For this reason, random graphs should be one of the most challenging sets for our method. Especially for random graphs, we do not expect that training examples generalize well to test examples as both sets could be very independent. Even given enough training examples, we expect there always will be graphs that do not share common properties with any other graph.

We simulated CQCNN's learning process for random graphs, each sampled uniformly from the set of all possible graphs with $n$ vertices and $m$ edges. The learning performance results are shown in Fig.~\ref{fig:randomLearning} for $n=15, 20,$ and $25$, $m$ is chosen uniformly from $n-1$ to $(n^2-n)/2$. In our simulations we observe that the loss after training is close to zero (below $3\times 10^{-3}$) for all these random graphs. 
In Fig.~\ref{fig:randomLearning}(a) we see that both recall and precision\footnote{Recall quantifies the fraction of correct predictions in a particular class, whereas precision identifies the fraction of a particular class predictions that turned out to be correct.} are $90\%$ for the ``classical'' part of the set, and is in the range of $25-35\%$ for the ``quantum'' part of the set. Overall, we see that our method made it possible to classify random graphs correctly much better than a random guess\footnote{The random guess will guess ``classical'' (or ``quantum'') class correctly in $50\%$ of the cases.} without performing any quantum walk dynamics simulations. Examples of correctly classified graphs are shown in Fig.~\ref{fig:randomLearning}(b)-(c).

\section*{Discussion} \label{sec:discussion}

Recently speedup problem extensively has been discussed in the framework of quantum computation purposed to accelerate the solution of familiar optimization problems by using quantum hardware~\cite{Kechedzhi2016,Albash2018}. However, predicting a quantum speedup in this hardware represents a complex problem that depends on many physical parameters including size and topology of the system~\cite{Alodjants2017,Lewenstein2017,Hamze2014,Smolin2014}. In this paper we proposed a new machine learning method to predict a speedup of quantum transport. This method is based on training a discriminative classifier, that is, a specially designed convolutional neural network (CQCNN). We have generated the training examples, each consisting of an adjacency matrix and a corresponding label (``classical'' or ``quantum''), by simulating the random walk dynamics of classical and quantum particles. The generated examples were used to train CQCNN with a stochastic gradient descent algorithm.

By training CQCNNs we demonstrated in Fig.~\ref{fig:linesLearning} that the neural network is able to learn to classify the quantum speedup, and to match the results obtained by our simulations. First, CQCNN learns to approximate given examples very well by representing the quantum and classical properties of graphs in its weights: CQCNN compresses up to $2268$ adjacency matrices with $49$ entries each\footnote{Which is the case for line graphs with $7$ vertices as the training set consisted of $90\%$ of $7!/2$ the total number of line graphs, see Fig.~\ref{fig:linesLearning}(b) for $n=7$.} into just $29$ real parameters. Second, CQCNN automatically learns what graph features are important for quantum speedup. We identified that for line graphs these correlations correspond to well-explainable graph properties. Additionally, the neural network learns that many graphs are isomorphic, with no indication of over-fitting on adjacency matrix features. Third, we demonstrated good generalization capacity of the constructed CNN. The neural network was correctly classifying not only previously unseen graphs of the same size, but also of sizes that were never given to train the network. 
For the line graphs of the same size the average accuracy was shown to be above $90\%$, and $60-85\%$ in the case of the larger graph sizes. We believe that this performance is strong as we know that test examples do not necessarily share any structural similarities with training examples.

Finally, the presented approach was applied to random graphs with up to $25$ vertices. Although the space of possible labeled graphs is more than $2^{200}$ graphs (see Ref.~\cite{slone1964online} for $25$ vertices), with only $1000$ randomly generated training examples we proved that it is possible to significantly improve over the random guess. We, however, believe that this classification performance can be further improved by using more training examples and by optimizing over CQCCN's hyperparameters.

%The generalization capacity of CQCNN was analyzed using a test set of graphs. For the line graphs with up to $7$ vertices the average accuracy was shown to be above $90\%$.
%We showed that it's also possible to correctly classify graphs even if these larger sizes were never presented to CQCNN before. In this case we have given examples of line graphs of size $4$ and $5$, and tested CQCNN on sizes of $6,7,8,9$, and $10$. Even though the graphs of these sizes were never given to the network, and there are probably several cases that are not generalizable, CQCNN showed the accuracy of $60-85\%$. 

The presented machine learning methodology can be used to find novel topologically large-scale graphs and circuits which exhibit maximal quantum speedup. At the same time our results might be specifically important in material science and biophotonics for a deeper understanding and designing of novel materials with unique quantum transport properties. 

%But is CQCNN is capable of generalizing beyond the seen results of our quantum and random walk simulations? To answer this question, we did the analysis of the neural network accuracy on unseen graphs. It turned out that the accuracy is above $90\%$ for the majority of studied sets of graphs.

\section*{Methods} \label{sec:methods}

In this section we give additional details on the machine learning methodology and the learning methods.

\subsection*{Quantum walks on graphs}\label{sec:QW_graphs}
In the following, we describe the quantum walk dynamics on graphs, and give more details on simulations that were performed in this paper.

We consider $n\times n$ adjacency matrices $A\in \mathcal{A}$ that describe undirected connected graphs $G\in \mathcal{G}$ with $n$ vertices on which classical and quantum walks are simulated. A graph $G$ is specified by the set of vertices $\mathcal{V} = \{1,\dots,n\}$ and the set of edges $\mathcal{E}$. All edges $(v, u) \in \mathcal{E}$ are described by a pair of vertices $v, u\in \mathcal{V}$.
As the graphs $G$ that we consider are undirected, $(u, v)= (v, u)$ and all matrices $A$ are symmetric: $A_{ij}=A_{ji}$.
Without the loss of generality, we label the vertices $v_\mathrm{i}=1$ and $v_\mathrm{t}=2$ as the ``initial'' and the ``target'' vertices. Given an adjacency matrix $A$, we simulate classical and quantum continuous-time walks during the time $t_\mathrm{max}$, which depends on the probability of detecting a particle. The results of the simulations are classical and quantum dependencies of the probability of detecting a particle in $v_\mathrm{t}$ at time $t\leq t_\mathrm{max}$. From these two dynamics we obtain the information about the time particle is in $v_\mathrm{t}$ with threshold probability $p_\mathrm{th}=1/\log n$. Given the two time values we can predict if there exists some quantum advantage of using a quantum particle for reaching $v_\mathrm{t}$ on a given graph.

The classical continuous-time random walk (CTRW) is simulated by solving the following differential equation
\begin{equation}\label{classicalWalk}
  \frac{\mathrm{d}\mathrm{p}(t)}{\mathrm{d}t} = (T-I) \mathrm{p}(t),
\end{equation}
where $\mathrm{p}(t)$ is a vector of probabilities $\mathrm{p}_v(t)$ of detecting a classical particle in vertices $v\in \mathcal{V}$ of the graph; $I$ is the identity matrix of size $n\times n$. The transition matrix $T$ is a matrix of probabilities $T_{vu}$ for a particle to jump from $u$ to $v$.
As we would like to ``catch'' the particle in $v_\mathrm{t}$, the edges $(v,v_\mathrm{t})$ that lead to $v_\mathrm{t}$ are made directed. This modification is implemented by introducing a new adjacency matrix $A^{c}$ which is equal to $A$ apart from the column $n$: $A^{c}_{nv}=0$, $\forall v\in \mathcal{V}\setminus n$, and $A^{c}_{nn}=1$.
The transition matrix can be obtained from the corresponding adjacency matrix $A^{c}$ by dividing all entries in a $v$-th column of $A^{c}$ by the in-degree of the vertex $v$, for all $v\in \mathcal{V}$.
This introduced modification of $A$ effectively makes the underlying graph $G^{c}$ directed such that a classical particle cannot escape $v_\mathrm{t}$ once it is there.

The solution of the differential equation in Eq.~(\ref{classicalWalk}) is
\begin{equation}\label{classicalWalkSolution}
  \mathrm{p}(t) = \mathrm{e}^{(T-I)t}\mathrm{p}(0) = \mathrm{e}^{-t}\mathrm{e}^{T t}\mathrm{p}(0),%\smallskip
\end{equation}
where $\mathrm{p}(0) = (1, 0, \dots, 0)^\mathrm{T}$ is a probability vector corresponding to a classical particle initially located in $v=1$. The dynamics in Eq.~(\ref{classicalWalkSolution}) is known as node-centric CTRW~\cite{aldous2002reversible,masuda2017random}. Node-centric CTRWs have a property that a particle leaves a vertex $v$ with the same rate for all vertices $u\in \mathcal{V}$. In the considered case the trajectories are statistically the same as those of the discrete-time random walk (DTRW), hense the dynamics of $\mathrm{p}(t)$ in Eq.~(\ref{classicalWalk}) can be viewed as a ``continuization'' of the DTRW dymanics.

The continuous-time quantum walk (CTQW) dynamics is simulated by solving the Gorini-Kossakowski-Sudarshan-Lindblad (GKSL) equation
\begin{equation}\label{quantumWalk}
  \frac{\mathrm{d}\rho(t)}{\mathrm{d}t} = - \frac{i}{\hbar}\left[\mathcal{H}, \rho(t)\right] + \gamma\left(L\rho(t)L^\dagger-\frac{1}{2}\{L^\dagger L, \rho(t)\}\right),
\end{equation}
with the Hamiltonian $\mathcal{H} = \hbar A^{q}$. $A^{q}$ is an adjacency matrix of size $(n+1)\times(n+1)$ and is equal to $A$ apart from adding an $(n+1)$-th row and an $(n+1)$-th column of zeros: $A_{v(n+1)}=A_{(n+1)v}=0, \forall v\in \mathcal{V}\cup\{n+1\}$. The new $A^{q}$ matrix corresponds to a graph $G^{q}$ with an additional ``sink'' vertex $v_\mathrm{sink}=n+1$. This sink vertex serves as an auxilary vertex where a quantum particle is kept captured once it ends there. The only way the particle can end there is by decaying from $v_\mathrm{t}$, this process is mathematically taken care of by the operator $L=\ket{n+1}\bra{n}$. Physically, $L$ introduces incoherence in the unitary CTQW dynamics described by $\mathcal{H}$, by moving the quantum particle from $v_\mathrm{t}$ to $v_\mathrm{sink}$ with rate $\gamma$. In general, the rate $\gamma$ dramatically influences the CTQW dynamics: if $\gamma=0$ -- the dynamics is coherent and we will never observe the particle in $v_\mathrm{sink}$, if the value of $\gamma$ is large (e.g., $10^5$) -- we might never observe the particle in $v_\mathrm{sink}$\footnote{This effect is known as the Zero effect, a vertex is measured to frequently so the particle never appears there.}. Because there is no universally best value for the $\gamma$ parameter for all graphs $G^{q}$, we use $\gamma=1$ throughout the paper.

We solve to the GKSL equation numerically with the initial condition $\rho(0)=\ket{1}\bra{1}$ and observe the dynamics of $\rho_{(n+1)(n+1)}(t)$ that is equal to the population in $v_\mathrm{sink}$ at time $t$. The function $\rho_{(n+1)(n+1)}(t)$ is a positive and an increasing function of time. Note that, opposite to the case of the CTRW, in the CTQW the probability of detecting the particle does not necessarily go to one with time.

We next compare $\mathrm{p^q}(t)\equiv\rho_{(n+1)(n+1)}(t)$ and $\mathrm{p^c}(t)\equiv\mathrm{p}_n(t)$ against $p_\mathrm{th}$. The time at which $\mathrm{p^q}(t)>p_\mathrm{th}$ or $\mathrm{p^c}(t)>p_\mathrm{th}$ is called the hitting time for quantum or classical particle, respectively.

\subsection*{Convolutional neural network architecture}
In this section we describe in detail how the convolutional neural network, which is used in this paper, is constructed.

\begin{figure}[t]
	\centering
	\includegraphics[width=1\linewidth]{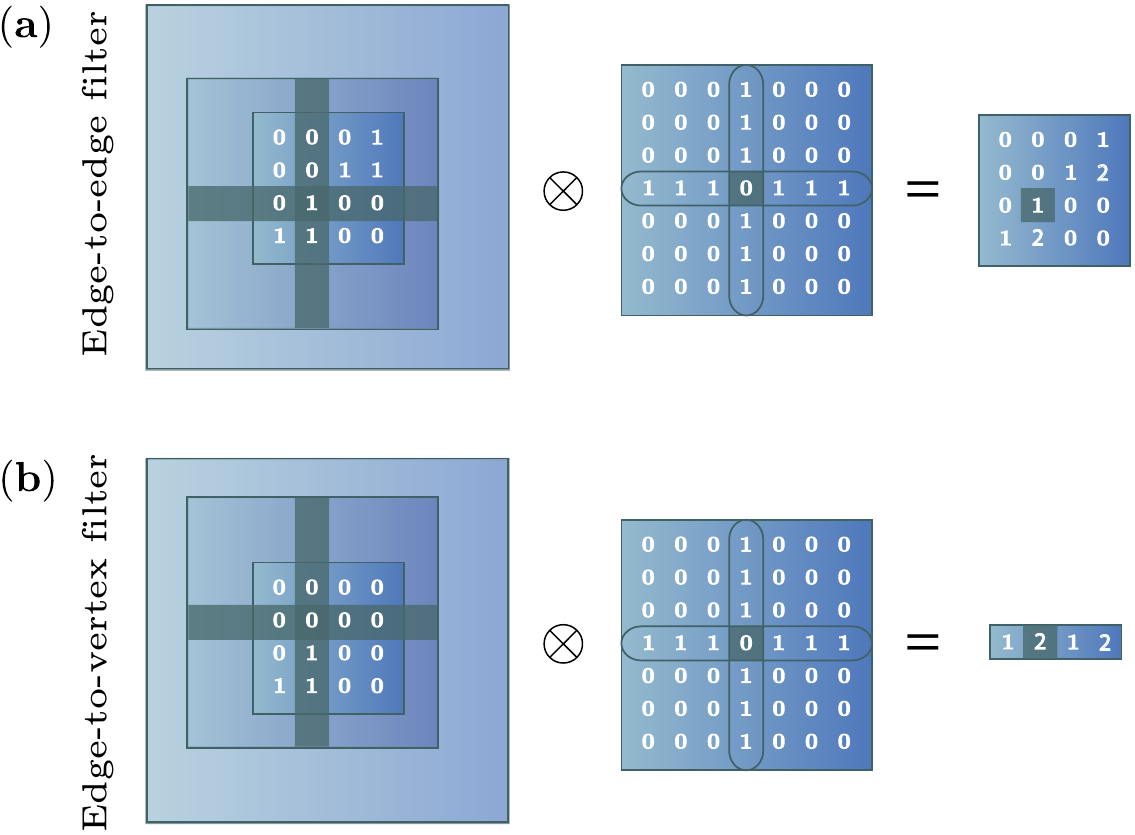}
	\caption{(a) The working principle of the edge-to-edge filter. An example of processing the adjacency matrix of a line graph with $4$ vertices is shown. A feature map shows that there are four edges with one neighboring edge, and two edges with two neighboring edges. (b) The working principle of the edge-to-vertex filter. An example of processing the adjacency matrix (already without its symmetric part) of a line graph with $4$ vertices is shown. A feature map shows that there are two vertices with one neighboring edge, and two vertices with two neighboring edges.}
	\label{fig:ETE}
\end{figure}

We are using a specifically designed convolutional neural network, CQCNN, to learn from different graphs. The architecture of this neural network is shown in Fig.~\ref{fig:schemes}(c). CQCNN, which we specifically designed to work with graphs, consists of a two-dimensional input layer that takes one graph represented by an adjacency matrix $A$. This layer is connected to several convolutional layers, the number of which depends on the number of vertices $n$ of the input graph. The first convolutional layer consists of six filters (or, feature detectors) that define three different ways of processing the input graph. These three ways are marked by different colors (green, red, blue) in Fig.~\ref{fig:schemes}(c). The weights and types of filters determine what specific features are detected. The first type of filters detects how well the $v_\mathrm{i}$ vertex is connected to the rest of the graph by extracting features from the $T^k$ matrices, where $k$ are integer numbers. The second type of convolutions detects the same, but for the $v_\mathrm{t}$ vertex. The third filter type looks at connectivities within the graph and detects how well each vertex is connected to other vertices. These three filter types are applied in several layers together with identity filters that propagate extracted features further. These layers are followed by a filter that deletes symmetric parts of all the matrices. It is done to eliminate redundant information, as all the matrices are still symmetric after being processed by all these fixed filters. At the next layer we apply $n$ filters of the fixed $3\times 3$ size with variable parameters in order to find relations between different edges. The last layer of filters summarizes all the information about the edges in the vertices description, by that decreasing the number of neuron values to a polynomially smaller number of next layer's neuron values.
The extracted features are next flattened and connected to two fully connected layers on neurons. Neurons in the first fully connected layer have a rectified linear unit (ReLU) activation function, which helps to construct a nonlinear function, and let the last layer map the learned features to $0$ or $1$ label (two output neurons in Fig.~\ref{fig:schemes}(c)).

CQCNN makes a choice between classical and quantum classes based on the values of two output neurons. The predicted class is defined as an index of a neuron with the largest output value:
\begin{equation}\label{predictedClass}
  \mathrm{class} = \argmax_m y(m).
\end{equation}
The network learns by stochastic gradient descent algorithm that takes the cross entropy loss function in Eq.~(\ref{lossFunction}).

The filters that we constructed in the described neural network architecture are essential to the success of learning. First, the edge-to-edge (ETE) filter allows the network to see how many neighboring edges each edge has. The process of obtaining a feature map from an input ``image'' using the edge-to-edge filter is shown in Fig.~\ref{fig:ETE}(a). Given an input matrix $A$ the ETE filter outputs the following matrix with components:
\begin{equation}\label{ETE}
  F^\mathrm{ETE}_{ij}(A) = \left[\sum_{k=1}^n \left(A_{ik} + A_{kj}\right) -2A_{ij} \right] A_{ij}
\end{equation}

\noindent The second important filter is the edge-to-vertex (ETV) filter. This filter allows summarizing information about the edges in the vertices. The filtering procedure takes an input matrix $A$ and outputs a vector with components:
\begin{equation}\label{ETN}
  F^\mathrm{ETV}_{i}(A) = \sum_{k=1}^n \left(A_{ik} + A_{ki}\right) - 2A_{ii}.
\end{equation}
The working principle of this filter is visualized in Fig.~\ref{fig:ETE}(b).

% \subsection*{Details of learning performance}
% Here we give more details on the performance of the neural networks in recognizing the difference between quantum and classical hitting times.

% \\\bigskip\bigskip\bigskip

\section*{Acknowledgment}
This work was financially supported by the Government of the Russian Federation, Grant 08-08, and by RFBR grants No. 19-52-52012 MHT-a and No. 17-07-00994-a.  
\bigskip\medskip

\section*{Data Availability}
The developed algorithms and the generated datasets are available from the corresponding author on a reasonable request.

\section*{Author Contributions}
A.A.M., L.E.F., and A.A. designed and performed research. A.A.M. performed the simulations and wrote the initial manuscript. A.A.M., L.E.F., and A.A. discussed the results and contributed to the manuscript.

\section*{Additional Information}
\textbf{Competing Interests:} The authors declare no competing financial or non-financial interests.

%\bibliographystyle{ieeetr}
% \bibliography{qwml.bib}

\end{document}